# The dynamics of the *min* proteins of *Escherichia coli* under the constant external fields

**Paisan Kanthang,[1] Waipot Ngamsaad,[1]Charin Modchang,[1] Wannapong Triampo,[1*] Narin Nuttawut,[1] I-Ming Tang,[1] Yongwimol Lenbury[2]**

[1]Department of Physics and Capability Building Unit in Nanoscience and Nanotechnology, Faculty of Science, Mahidol University, Bangkok 10400, Thailand

[2]Department of Mathematics, Faculty of Science, Mahidol University, Bangkok 10400, Thailand

## Abstract

In *E. coli* the determination of the middle of the cell and the proper placement of the septum is essential to the division of the cell. This step depends on the proteins MinC, MinD, and MinE. Exposure to a constant external field e.g., an electric field or magnetic field may cause the bacteria cell division mechanism to change resulting in an abnormal cytokinesis. To have insight into the effects of an external field on this process, we model the process using a set of the deterministic reaction diffusion equations, which incorporate the influence of an external field, *min* protein reactions, and diffusion of all species. Using the numerical method, we have found some changes in the dynamics of the oscillations of the *min* proteins from pole to pole when compared that of without the external field. The results show some interesting effects, which are qualitatively in good agreement with some experimental results.

*Key words:* external fields, *E. coli,* cell division, *min* proteins, MinCDE oscillation

*Corresponding author, E-mail: **wtriampo@yahoo.com**

## I. INTRODUCTION

In nature, the cell division or cytokinesis process is a crucial event in the life of every organism. Most bacteria divide symmetrically in a process that is subject to extensive regulation to ensure that both newly formed daughter cells contain a copy of the chromosome. Symmetric division seems very simple yet is poorly understood on a molecular biological level. For *Escherichia coli* and other rod-like bacteria, evidences have accumulated over the past few years which indicate that the separation into two daughter cells is achieved by forming a septum perpendicular to parent cell's long axis. To induce the separation, the FtsZ ring (Z ring), a tubulin-like GTPase is believed to initiate and guide the septa growth by a process called contraction (Lutkenhaus 1993). The Z ring is usually positioned close to the center, but it can also form in the vicinity of the cell poles. Two processes are known to regulate the placement of the division site: nucleoid occlusion (Woldringh 1991) and the action of the *min* proteins (de Boer 1989). Both processes interfere with the formation of the Z ring that determines the division site. Nucleoid occlusion is based on cytological evidence that indicates that the Z ring assembles preferentially on those portions of the membrane that do not directly surround the dense nucleoid mass (Mulder 1989).

The *min* proteins that control the placement of the division site are the MinC, the MinD, and the MinE proteins (de Boer 1989). Experiments, involving the use of modified proteins show that MinC is able to inhibit the formation of the FtsZ-ring (de Boer 1990). MinD is an ATPase that is connected peripherally to the cytoplasmic membrane. It can bind to the MinC and activate the function of the MinC (de Boer 1991, Huang 1996). Recent studies show that the MinD can also recruit the MinC to the membrane. This suggests that the MinD stimulates the MinC by concentrating the MinC near to its presumed site of activation (Hu 1999, Raskin 1999a). MinE provides topological specificity to the division inhibitor (Fu, 2001). Its expression results in a site-specific suppression of the MinC/MinD action so that the FtsZ assembly is allowed at the middle of the cell but is blocked at the other sites (de Boer 1989). In the absence of the MinE, the MinC/MinD is distributed homogeneously over the entire membrane. This results in a complete blockage of the Z-ring formation. The long filamentous cells, which are subsequently formed would not be able divide (Hu 1999, Raskin 1999a, Raskin 1999b, Rowland 2000). Using fluorescent labeling, the MinE was shown to attach to the cell wall only in the presence of the MinD (Huang 2003, Raskin 1997). As MinD dictates the location of MinC, the latter would oscillate by itself. This would result in the concentration of the division inhibitor at the membrane on either cell end, alternating between being high or very low every other 20 s or so (Hu 1999, Raskin 1999a). The presence of MinE is not only required for the MinC/MinD oscillation, it is also involved in setting the frequency of the oscillation cycle (Raskin 1999b). Several sets of evidence indicate that the MinE localization cycle is tightly coupled to the oscillation cycle of MinD. Recent microscopy of the fluorescent labeled proteins involved in the regulation of *E. coli* division have uncovered stable and coherent oscillations (both spatial and temporal) of these three proteins (Hale 2001). The proteins oscillate from one end to the other end of the bacterium, moving between the cytoplasmic membrane and cytoplasm. The detail mechanism by which these proteins determine the correct position of the division plane is currently unknown, but the observed pole-to-pole oscillations of the corresponding distribution are thought to be of functional importance. Under different culture conditions and/or environment changes, (e.g., pH, light, and external field) the pole-to-pole oscillations could affect the growth of the bacteria. Here we discuss only the effects of a magnetic field.

Magnetic fields (MF) can affect various biological functions of living organisms, e.g., DNA synthesis and transcription (Phillips 1992), as well as ion transportation through cell membranes (Liburdy 1993). Almost all living organism experience magnetic fields arising from one or another sources. The strength of the geomagnetic field on the surface of the earth is approximately 0.50-0.75 gauss. There have been several studies over the past decades on the effects of exposure to the magnetic field and several of them have produced conflicting results. The growth rate of the Burgundy wine yeast is seen to decrease when an extremely low magnetic flux density (MFD) of 4 gauss is applied (Kimball 1938). However, the growth of *Trichomonas vaginalis* is accelerated when it is exposed to 460-1200 gauss (Genkov 1974). The growth of *Bacillus subtilis* increases when exposed to 150 gauss and decreases when exposed to more than 300 gauss (Moore 1979). Similar results are reported for *Chlorella*, an exposure of less than 400 gauss increases the growth, while exposure to 580 gauss decreases the growth (Takahashi 1985). Most studies point to the magnetic field influencing the growth and survival of the living organisms differently at different MFD (Yamaoka 1992, Singh 1994, Tsuchiya 1996, Horiuchi 2001, Piatti 2002). Researchers have also studied the effects of the magnetic fields on the bacteria at the enzyme level (Utsunomiya 2003) and at the genetic level (Horiuchi 2001).

In the present work, we use a novel approach to investigate the influence of the constant external fields on the cytokinesis mediated by *min* protein pole-to-pole oscillation. We propose a mathematical model and then numerically solve it to study how the *min* protein oscillation mechanism for the bacteria cell division may change.

## II. Model

The most natural process can be described by the reaction-diffusion equations which have often been used in biological applications to model self-organization and pattern formation (Nicolis 1977). These mathematical models have two sub component processes. The first process is the reaction process which represents the kinetic behavior with the given time such as the self-organization of the biological system. The second process is the diffusion process which represents the random motion of the system. For the molecular level, the diffusion process due to the random motion of molecules in a medium.

For our work, we focus on the reaction-diffusion model for describe the dynamic of the *min* proteins on the *E. coli* bacteria. This bacterium have the size which approximately $2-6\mu m$ in length and around $1-1.5\mu m$ in diameter. The *E. coli* have division roughly every hour via cytokinesis. From the dynamic model of *E. coli* cell division proposed by Howard *et. al.* which have not representing the external field. We will start with the set of one dimensional deterministic coupled reaction-diffusion equations for describe the dynamics of MinD and MinE. For the effect of the external field on the *min* proteins, we represent in the term of the external field parameter multiply by the flux of *min* proteins which mean that the external field can influence the change in the concentrations (Zemskov 2003, Munuzuri 1995) of the *min* proteins at each position in *E. coli* cell and at any given time. The dynamic model for describe the *min* proteins under the external field, written as:

$$\frac{\partial C_D}{\partial t} = D_D \frac{\partial^2 C_D}{\partial x^2} + J_D \frac{\partial C_D}{\partial x} - \frac{\sigma_1 C_D}{1+\sigma_1' C_e} + \sigma_2 C_e C_d \ , \qquad (1)$$

$$\frac{\partial C_d}{\partial t} = D_d \frac{\partial^2 C_d}{\partial x^2} + J_d \frac{\partial C_d}{\partial x} + \frac{\sigma_1 C_D}{1+\sigma_1' C_e} - \sigma_2 C_e C_d \ , \qquad (2)$$

$$\frac{\partial C_E}{\partial t} = D_E \frac{\partial^2 C_E}{\partial x^2} + J_E \frac{\partial C_E}{\partial x} - \sigma_3 C_D C_E + \frac{\sigma_4 C_e}{1+\sigma_4' C_D} \ , \qquad (3)$$

$$\frac{\partial C_e}{\partial t} = D_e \frac{\partial^2 C_e}{\partial x^2} + J_e \frac{\partial C_e}{\partial x} + \sigma_3 C_D C_E - \frac{\sigma_4 C_e}{1+\sigma_4' C_D} \qquad (4)$$

where $C_D, C_E$ are the concentrations of protein MinD and MinE in the cytoplasm. $C_d, C_e$ are the concentrations of protein MinD and MinE on the cytoplasmic membrane. These parameters depend on the time $t$ and the position $x$. For $J_D$, $J_d$, $J_E$ and $J_e$ are the external field parameter of each species. These equations represent the time rate of change of the *min* protein concentrations in cytoplasm and on cytoplasmic membrane. Since the experimental results given in (Raskin 1999a), show that the MinC dynamics are similar to those of the MinD, we have not written out the equations for the MinC.

In this paper, we assume that the *min* protein molecules moving in the region of the external field which respect to the external field parameters of each species.

The normal form of the external field parameter is $J_i = \mu_i E$ (Zemskov 2003, Munuzuri 1995) where $E$ is the field strength and $\mu_i$ is the ionic mobility which index $i$ represent the species of molecule or chemical substance (for this paper $i = \{D, d, E, e\}$ ).

In this paper, we assume that the diffusion coefficients $(D_D, D_d, D_E, D_e)$ are isotropic and independent of $x$. The constant $\sigma_1$ represents the association of MinD to the membrane wall (Takahashi 1985). $\sigma_1'$ corresponds to the membrane-bound MinE suppressing the recruitment of MinD from the cytoplasm. $\sigma_2$ reflect the rate that the MinE on the membrane drives the MinD on the membrane into the cytoplasm. Based on the evidence of the cytoplasmic interaction between MinD and MinE (Huang 1996), we let $\sigma_3$ be the rate that cytoplasmic MinD recruits the cytoplasmic MinE to the membrane while $\sigma_4$ corresponds to the rate of dissociation of MinE from the membrane to the cytoplasm. Finally, $\sigma_4'$ corresponds to the cytoplasmic MinD suppressing the release of the membrane-bound MinE.

The novel feature of our model is the second terms on the right hand side. They represent the effect of the external field in the reaction-diffusion equation (Zemskov 2003, Munuzuri 1995) controlled by the external field parameter. These terms reflect the fact that the external field can influence the change in the concentrations of *min* proteins at each position in *E. coli* cell and at any given time. If the field parameter is very large, the dynamics of Min system would then be totally controlled by the external field.

In the present work, we consider an external field of constant strength. We assume that the *min* proteins can bind/unbind from the membrane and that the protein does not degraded during the process. The total amounts of each type of *min* proteins are conserved. The zero flux boundary conditions are imposed. This boundary condition gives a closed system.

## III. Numerical  results and discussion

We consider the situation where the external fields have the influence throughout the cell including the cytoplasm and the cell membrane. We have numerically solved the one-dimensional coarse-grained equations (1)-(4). We assume that there is homogeneity in all the spatial directions except for the x-direction. The explicit Euler method of integration (Press 2002) is used to obtain the numerical results. The size of *E. coli* is $2\mu m$ in length. We assume that the total time needed in the simulation will be approximately $10^4$ s. We discretized the space and time by taking the following increments to be $dx = 8 \times 10^{-3} \mu m$ and $dt = 1 \times 10^{-5} s$. (Howard 2001). The space covering the bacteria will be divided into 250 grid points, while the time will be divided into $10^9$ iteration steps. We use the uniform random initial condition with the number of *min* molecules in each cell being 3000 for the MinD population (de Boer, 1991) and 170 for the MinE population (Zhao 1995). For the boundary condition, we set the condition that the gradient of concentration is equal to zero at both ends of the bacterium. In our work, the *min* proteins have two species such as MinD and MinE. We assume the external field parameter $J_i$ which is proportional to the diffusion coefficient $D_i$. We also assume that the MinD and MinE have the same type of charges, so $J_D = J_E = J$ for *min* proteins in cytoplasm. For the *min* proteins on the membrane controlled by the characteristic properties in each species which represented by the ratio of diffusion coefficient such as $D_d / D_D$ for MinD and $D_e / D_E$ for MinE. Then the external field parameter of *min* proteins on the membrane represented by $J_d = (D_d / D_D)J$ for MinD and $J_e = (D_e / D_E)J$ for MinE. The external field parameter, we assume it to be of a constant value ranging from $J = 0$ to $J = 0.1$. For the values of the other parameters are as (Howard 2001, Howard 2003):

$D_D = 0.28\mu m^2 s^{-1}$, $\quad D_d = 0.003\mu m^2 s^{-1}$, $\quad D_E = 0.6\mu m^2 s^{-1}$, $\quad D_e = 0.006\mu m^2 s^{-1}$, $\sigma_1 = 20 s^{-1}$, $\quad \sigma'_1 = 0.028\mu m$, $\quad \sigma_2 = 0.0063\mu m s^{-1}$, $\quad \sigma_3 = 0.04\mu m s^{-1}$, $\quad \sigma_4 = 0.8 s^{-1}$, and $\sigma'_4 = 0.027\mu m$. In our analyses of the numerical results, we look at the time averaged values of the concentrations. Our focus in particular will be centered on the oscillation patterns which occur for different values of the external field parameter.

From our numerical results of the set of reaction-diffusion equations for non-external field which show that the most of the proteins will be concentrated at the membrane. This behavior means that the total dynamic of the Min system will be characterized by the dynamics of the *min* proteins on or near the membrane. Our numerical solutions show that the behavior of the Min system acted on by a constant field will depend on the strengths of the external field parameter $(J)$.

In Figure (1), we show that the space-time plots which represent the movement of the *min* proteins from pole to pole. When applied the strengths of the external field by increasing the external field parameters $(J)$. We see that the strengths of concentration of the *min* proteins increase at the left pole explicitly. This behavior mean that the Min system controlled by the strengths of the external field.

In Figure (2.1a-2.3a) and (2.1b-2.3b) show that the characteristic wave pattern of the *min* proteins at the left pole the midcell and the right pole. We have using the Fast Fourier Transformation (FFT) for measurement the period of the *min* protein waves which approximate to 128s for the left and the right pole, and 64s for the midcell. These results mean that the midcell has the height frequency when compares with the two poles. Which the midcell has the dynamic of the *min* proteins more than other regions. When we increase the external field parameters from $J = 0$ to $J = 0.1$, measure the period of these *min* protein waves which does not change. But, the amplitudes or the height concentrations are increase for the left pole and the midcell, and decrease for the right pole. The interesting of these figures is the wave pattern which does not change explicitly for the left and the right pole, but it changes explicitly for the midcell which mean that the dynamics of the *min* proteins at the midcell have sensitive to the external field because the period and the amount of *min* proteins is small value when compares with the other regions. If we compare figure (2.1a-2.3a) and (2.1b-2.3b) at each time , we see that the basis behavior of MinD and MinE does not change with the strength of external field which the time of MinD increase then MinE will decrease.

In Figure (3a) and (3b), it shows the time average of MinD and MinE concentration at the difference position within *E. coli* cell. When the parameter increase from $J = 0$ to $J = 0.1$ , the minimum of the time average curve of MinD have decreasing and shifting from the midcell to the right pole which mean that the MinE does not prevent MinD at the midcell. In nature, the MinE protein looks like a ring structure near the midcell that effectively positions the anti-MinCD activity to shield the midcell site from MinCD (Raskin 1997, Raskin 1999b). For MinE, the maximum of time average have shifting from the midcell to the right pole. These changing of the time average curves of *min* proteins confirm the external fields have influence with the dynamic of the Min system. However, the basis behavior of MinD and MinE does not change with the increasing of the external field parameters which is the regions at the concentration of MinD decrease then MinE increase. This behavior explicitly shows at the right pole of MinD and MinE of figure (3a) and (3b).

In Figure (4a) and (4b), we have measured the percentage of shifting of MinD and MinE from midcell for $J = 0$ to $J = 0.1$. We see that the trending of shifting of both min proteins will be straight line. We have also carried out a least square fit of the data to y = Bx + A. In Fig. 4a, we see that the shifts of the MinD concentrations increase as the field parameter *J* increase. The values of A and B in the linear dependence are A = 0.07619 and B = 65.14286 with a R = 0.99552. The same fit was done for the MinE concentrations. The results are shown in Fig 4b. The fitting parameter for the MinE proteins are A = -4.44089E-16 and B =40 and an R value of 1.0.

For the external field has a negative sense of direction, the results are very similar to those of positive external field as expected. Now however, the space-time plots show the increasing of *min* proteins at the right pole and the curve for the time averages of the concentration of the *min* proteins shifts from the midcell towards the left pole.

## IV. CONCLUDING REMARKS

According to the some current theories for the division process of the bacteria is required the accuracy of division site (de Boer, 1989) which it connected to the rapid pole-to-pole oscillations of the *min* proteins (Hu 1999, Raskin 1999b, Meinhardt 2001). Using a mathematical model to describe the dynamics of the *min* pole-to-pole oscillations, Howard *et al.* (2001), found that the midcell position in the *Escherichia coli* bacteria, correspond to the point where the time averaged MinD and MinE concentration were maximum and minimum, respectively. They also found that the concentrations of these two proteins were symmetric about the midcell position.

To see the influence of exposing the *E. coli* bacteria to an external field, we have added some additional terms to the reaction diffusion equations for the pole-to-pole oscillation of the *min* proteins in the *E. coli* bacteria proposed by Howard *et al.* The additional terms are the gradient terms appearing in eqns. (1)–(4). We then developed a numerical scheme to solve the resulting coarse-grained coupled reaction-diffusion equations. The results are shown in Figures 1 to 4. Our results show deviations from the results obtained by Howard *et al*., e.g. the concentrations of the MinD and MinE are no longer symmetric about the middle of the long axis nor are the maximum and minimum of the MinD and MinE concentrations at the middle of long axis. The shifting of the time average concentration of *min* proteins from the midcell appears to be dependent on the strength of the external field which these results represent the shifting of the division site. If the division process of a cell division point at the midcell, we have shown that an external field can interfere with the division process.

## Acknowledgements

We thank M. Howard, J. Wong-ekkabut and M. Chooduang for their useful comments and suggestion. This research is supported in part by Thailand Research Fund through the grant number TRG4580090 and RTA4580005. The IRPUS Program 2547 to Charin Modchang and W. Triampo is acknowledged. The Development and Promotion of Science and Technology talents program to Waipot Ngamsaad .

**Figures**

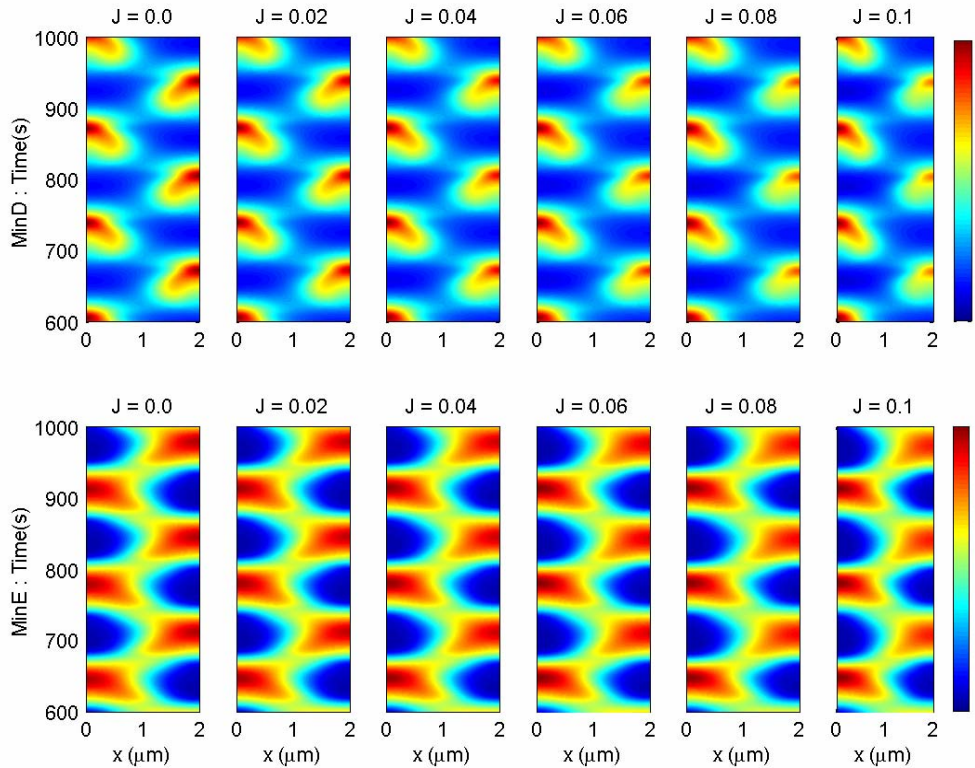

FIG. 1.  Space-time plots of the total MinD (above) and MinE (below) concentrations for various external field parameters of $J = 0$ to $J = 0.1$.  The color scale (blue to red) denotes an increase in the concentration from the lowest to the highest. The MinD depletion from midcell and the MinE enhancement at the midcell are immediately evident.  The times increase from bottom to top. The concentrations of *min* proteins shift to the left pole when the external field parameter is increased. The vertical scale spans a time passage of 1000 second. The horizontal scale spans the bacterial length ($2 \mu m.$).

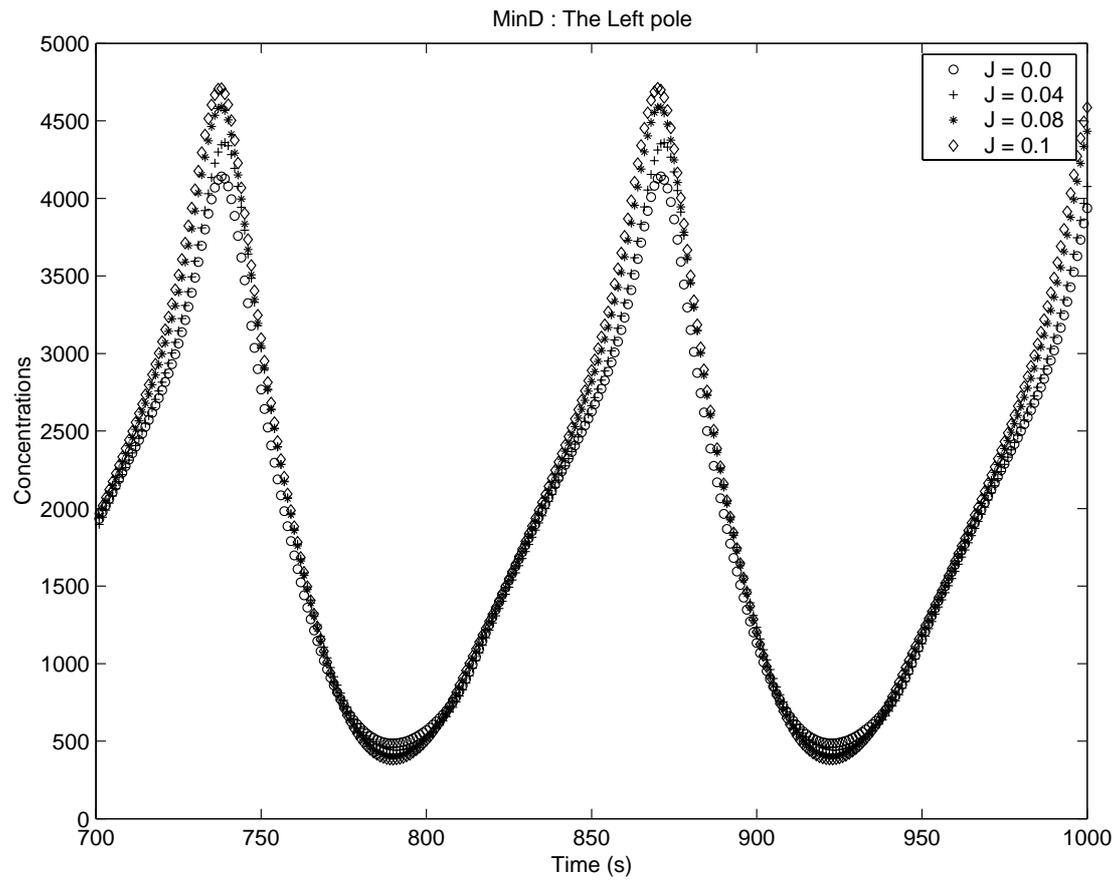

FIG. 2.1a. Show the oscillation pattern of MinD at the left pole for J=0.0, J=0.04, J=0.08 and J=0.1. The verticals scale span for the concentrations and the horizontal scale spans time in seconds.

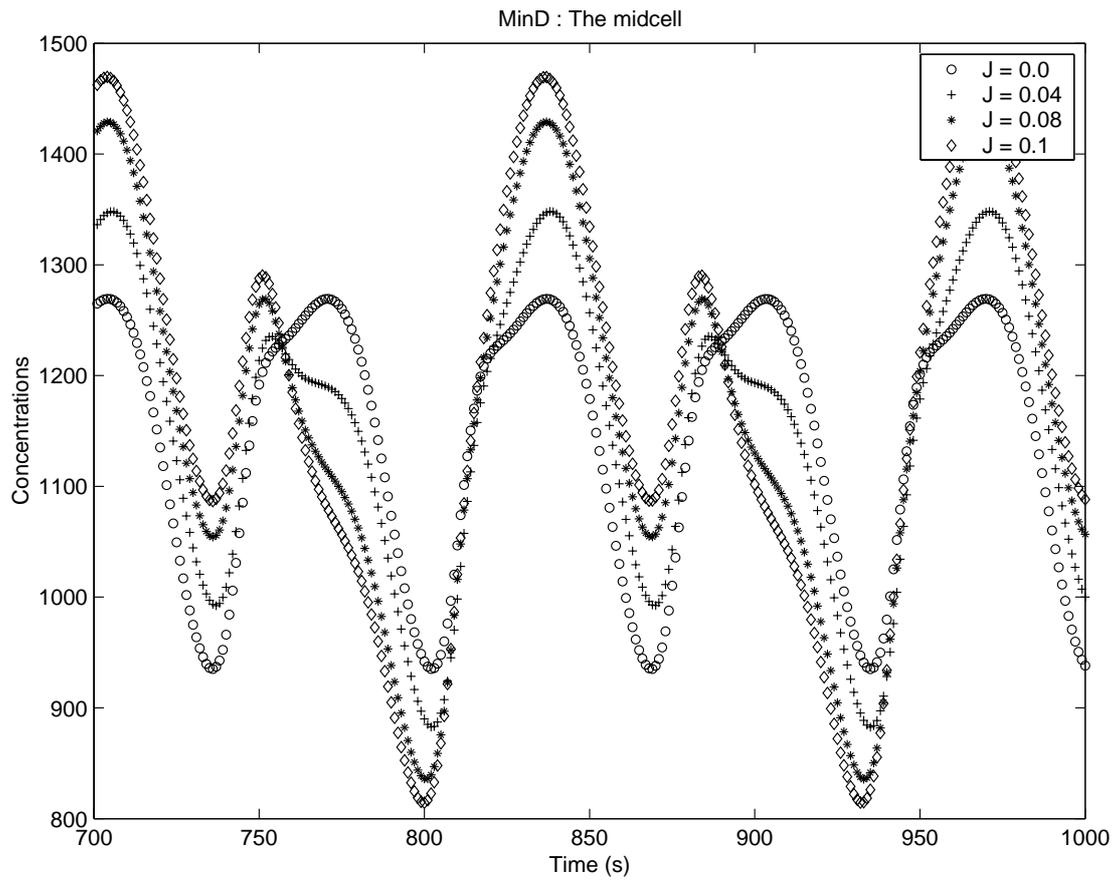

FIG. 2.2a. Show the oscillation pattern of MinD at the midcell for J=0.0, J=0.04, J=0.08 and J=0.1. The verticals scale span for the concentrations and the horizontal scale spans time in seconds.

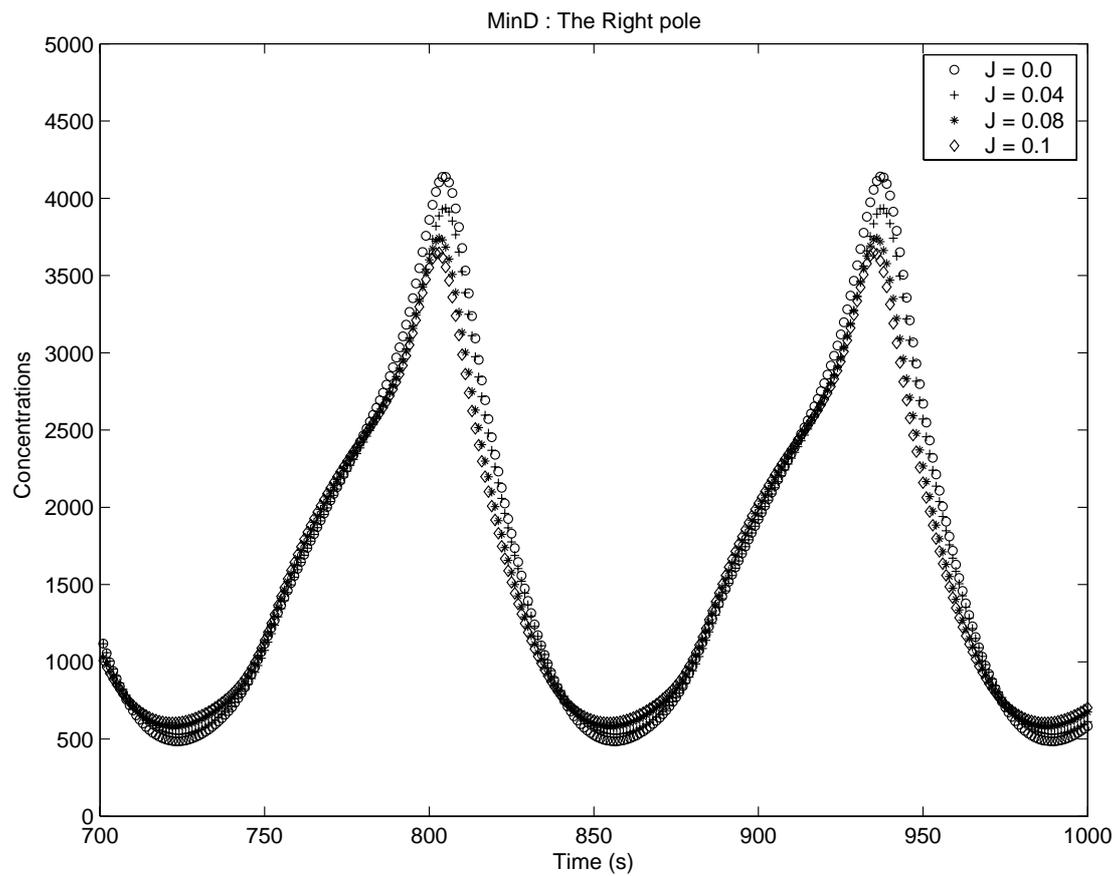

FIG. 2.3a. Show the oscillation pattern of MinD at the right pole for J=0.0, J=0.04, J=0.08 and J=0.1. The verticals scale span for the concentrations and the horizontal scale spans time in seconds.

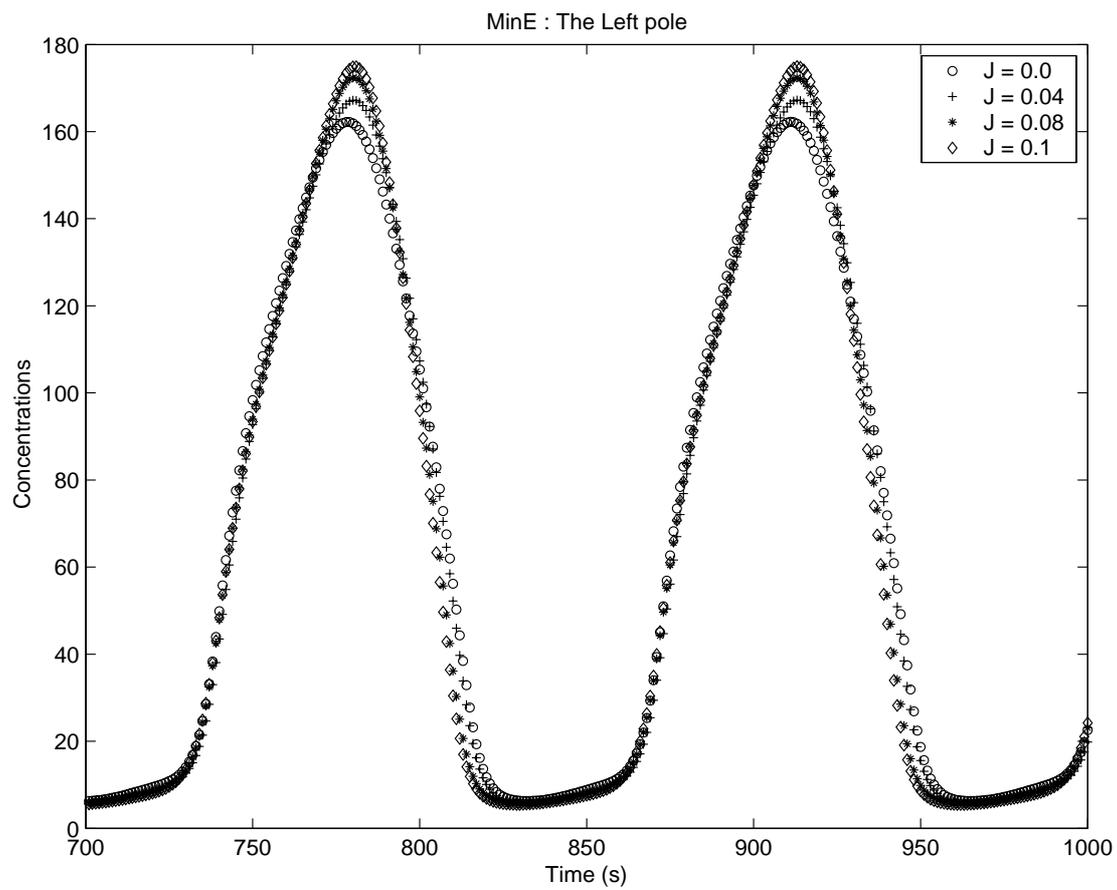

FIG. 2.1b. Show the oscillation pattern of MinE at the left pole for J=0.0, J=0.04, J=0.08 and J=0.1. The verticals scale span for the concentrations and the horizontal scale spans time in seconds.

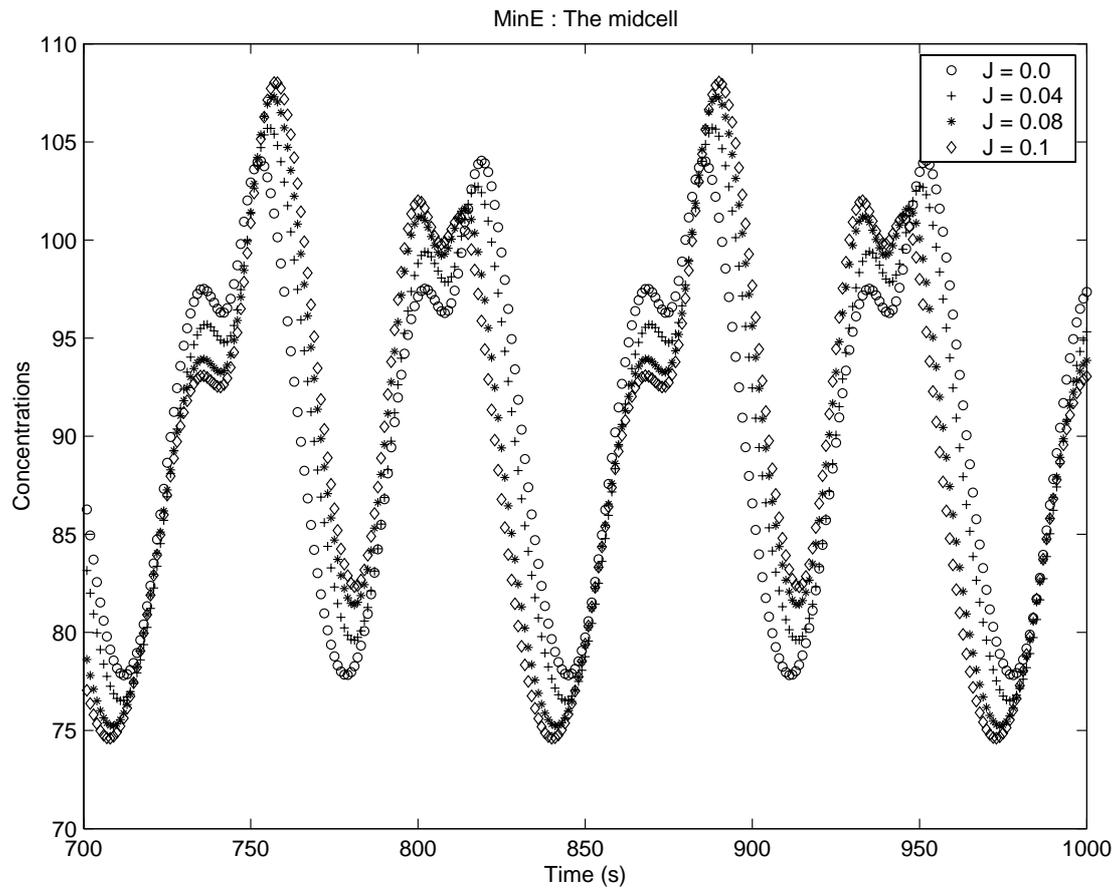

FIG. 2.2b. Show the oscillation pattern of MinE at the midcell for J=0.0, J=0.04, J=0.08 and J=0.1. The verticals scale span for the concentrations and the horizontal scale spans time in seconds.

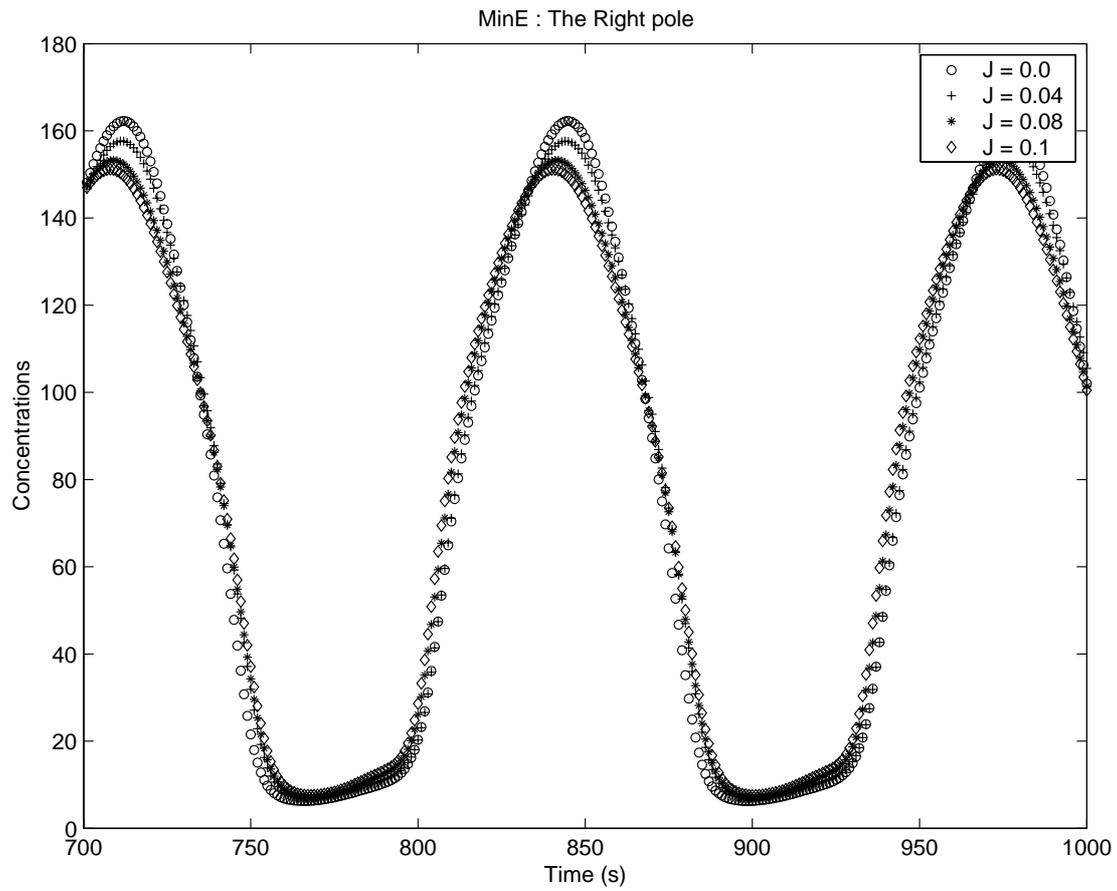

FIG. 2.3b. Show the oscillation pattern of MinE at the right pole for J=0.0, J=0.04, J=0.08 and J=0.1. The verticals scale span for the concentrations and the horizontal scale spans time in seconds.

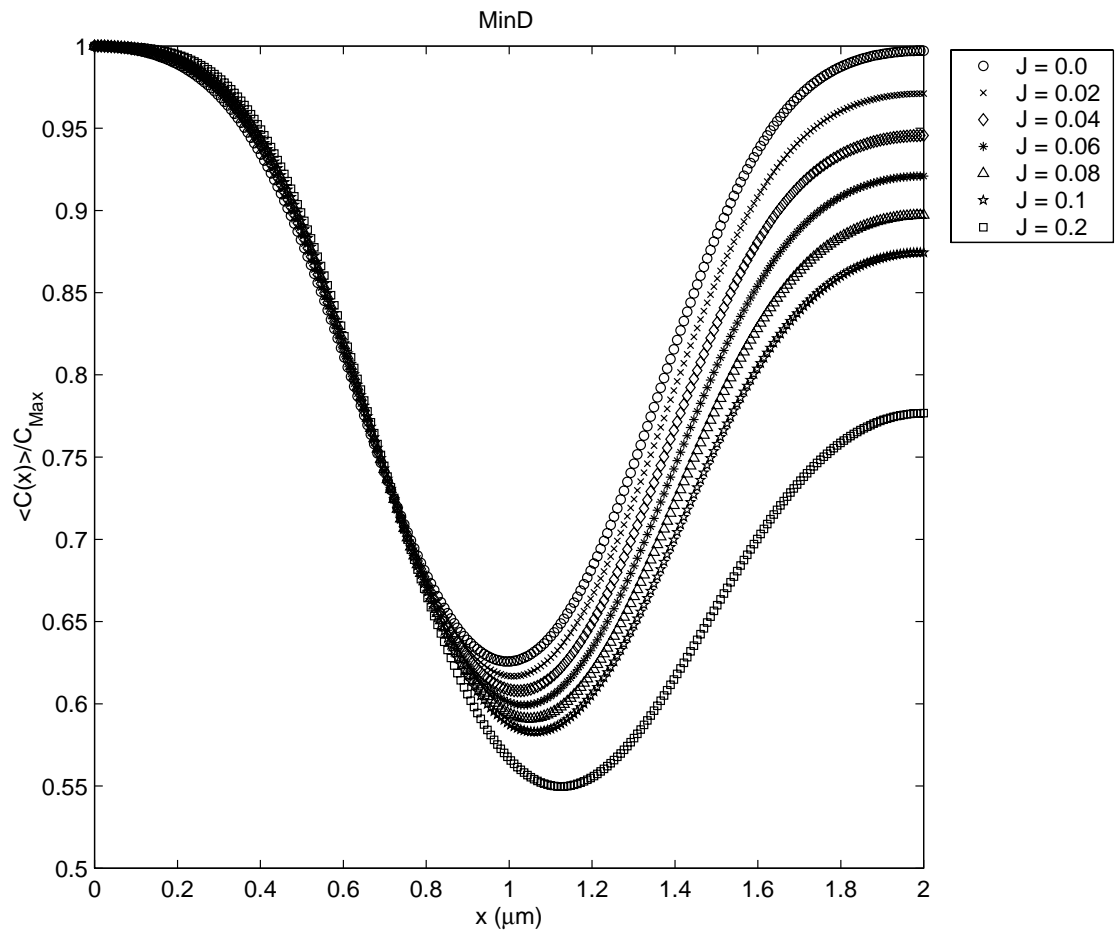

FIG. 3a. The time averages of the MinD concentrations $\langle C(x) \rangle / C_{\max}$, relative to their respective time-averaged maxima, as a function of position $x$ (in $\mu m$) along the bacterium axis under the influence of positive values of the static external field. Compared to the zero field case, the curves show a shift in the concentration of the MinD from the midcell depending on the strength of the field.

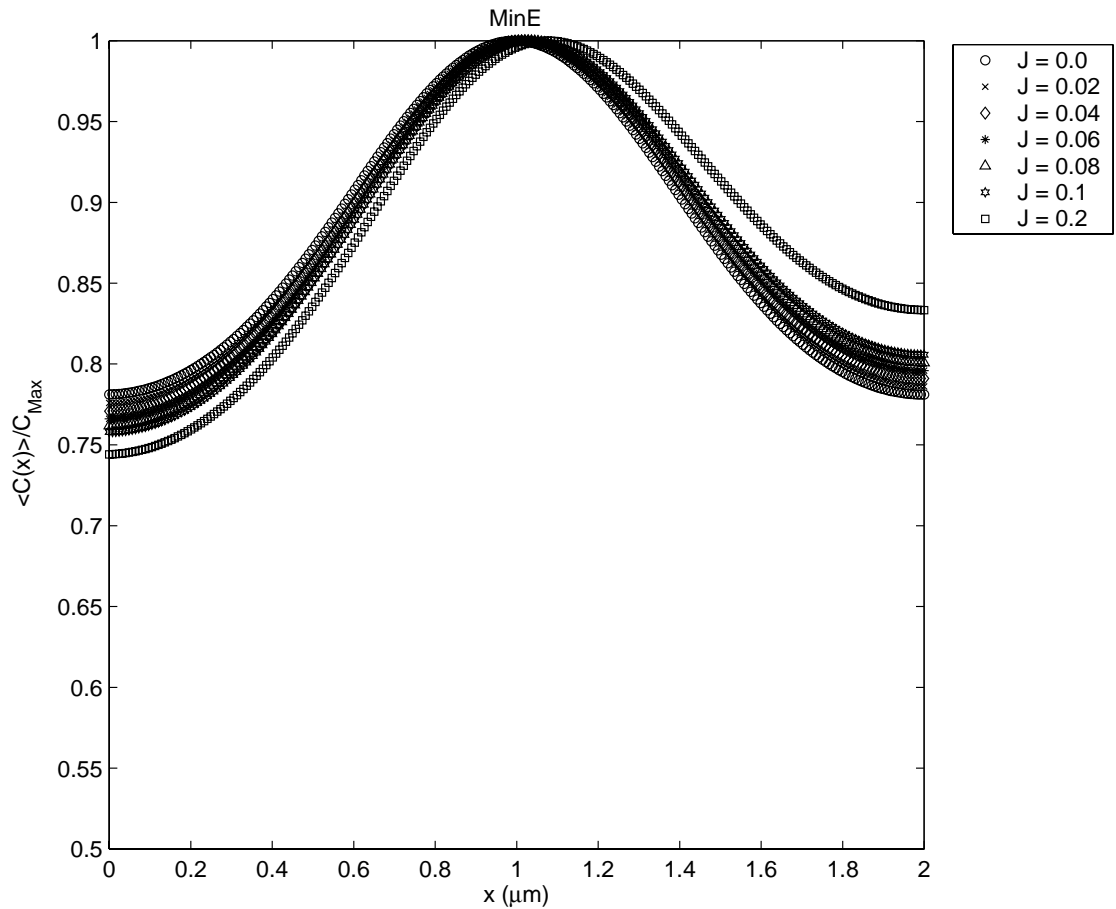

FIG. 3b. The time averages of the MinE concentrations $\langle C(x)\rangle / C_{max}$, relative to their respective time-averaged maxima, as a function of position $x$ (in $\mu m$) along the bacterium axis under the influence of positive values of the static external field. Compared to the zero field case, the curves show a shift in the concentration of the MinE from the midcell depending on the strength of the field. For the dynamics of MinE at the two poles when the external field parameters increase from J = 0 to 0.1, the time average values have decreasing at the left pole and increasing at the right pole.

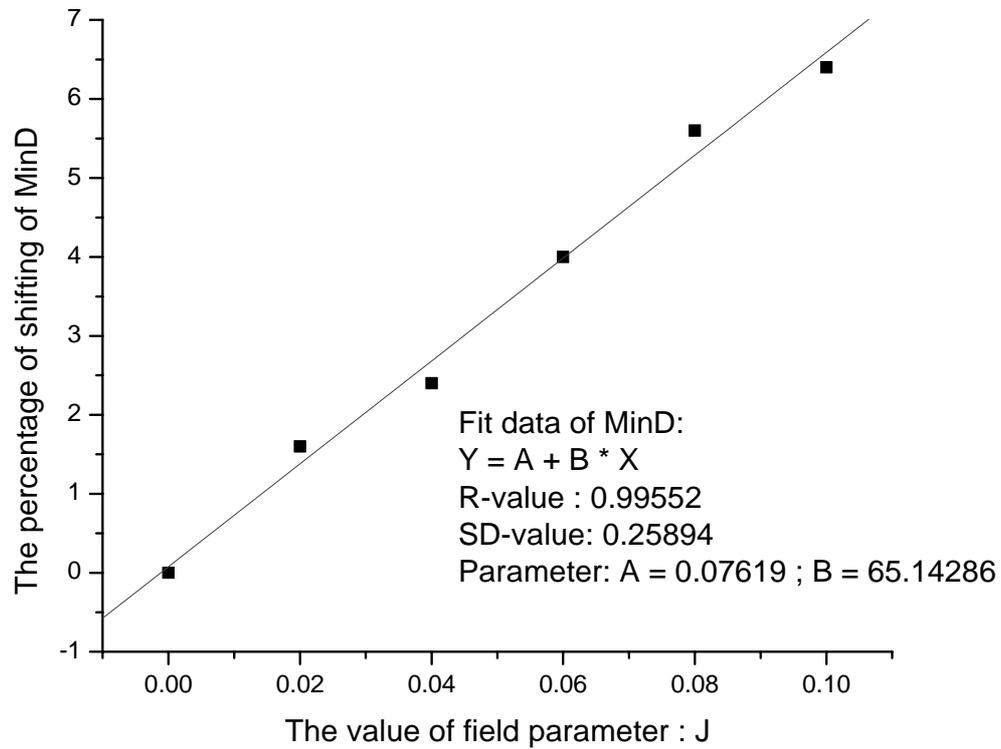

FIG. 4a. The shifting of the MinD from the midcell for positive values of the external field parameters: J=0.0, J =0.02, J=0.04, J=0.06, J=0.08 and J=0.1. The R-value is equal to 0.99552. The solid line shows the linear relation between the percentage of shifting of the MinD and the external field parameters.

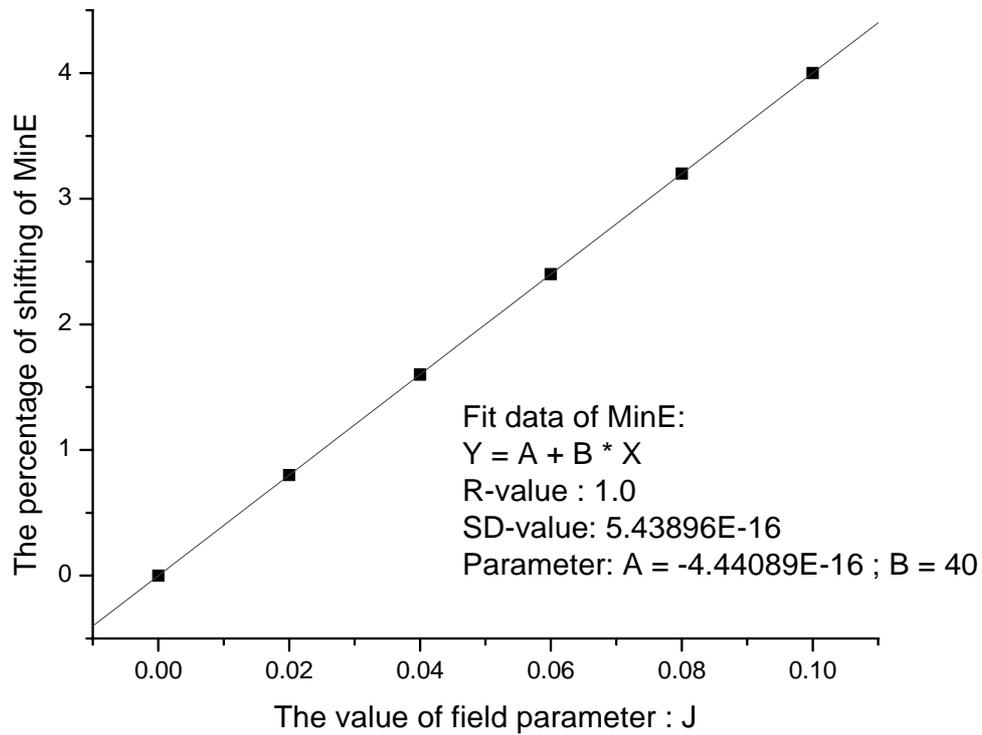

FIG. 4b. The percentage shifting of the MinE protein from the midcell position for positive values of the external field parameters: J=0.0, J =0.02, J=0.04, J=0.06, J=0.08 and J=0.1. The R-value is 1.0. The solid line is the best fit of the data to linear dependence of the shifting of the MinE concentrations on the external field parameters.